\documentclass[sigconf,nonacm]{acmart}

\usepackage{tcolorbox}
\usepackage{tabularx}
\usepackage{multirow}
\usepackage{array}
\usepackage{ragged2e}

\newcolumntype{L}{>{\raggedright\arraybackslash}X}

\AtBeginDocument{%
  }

\setcopyright{acmlicensed}
\copyrightyear{2024}
\acmYear{2024}
\acmDOI{XXXXXXX.XXXXXXX}

\setcopyright{none}
\definecolor{green}{HTML}{44AA99}
\definecolor{yellow}{HTML}{DDCC77}
\definecolor{blue}{HTML}{88CCEE}
\definecolor{red}{HTML}{CC6677}
\definecolor{darkred}{HTML}{DC3220}

\begin{document}

\title{A Longitudinal Analysis of Racial and Gender Bias in New York Times and Fox News Images and Articles}


\author{Hazem Ibrahim}
\authornote{Both authors contributed equally to this research.}
\email{hazem.ibrahim@nyu.edu}
\affiliation{%
  \institution{NYU Abu Dhabi}
  \city{Abu Dhabi}
  \country{UAE}
}

\author{Nouar AlDahoul}
\authornotemark[1]
\email{naa9497@nyu.edu}
\affiliation{%
  \institution{NYU Abu Dhabi}
  \city{Abu Dhabi}
  \country{UAE}
}

\author{Syed Mustafa Ali Abbasi}
\email{mustafaaliabbasi45@gmail.com}
\affiliation{%
  \institution{Lahore University of Management Sciences, Lahore, Pakistan}
  \city{}
  \country{}
}

\author{Fareed Zaffar}
\email{fareed.zaffar@lums.edu.pk}
\affiliation{%
  \institution{Lahore University of Management Sciences, Lahore, Pakistan}
  \city{}
  \country{}
}

\author{Talal Rahwan}
\email{talal.rahwan@nyu.edu}
\affiliation{%
  \institution{NYU Abu Dhabi}
  \city{Abu Dhabi}
  \country{UAE}
}

\author{Yasir Zaki}
\email{yasir.zaki@nyu.edu}
\affiliation{%
  \institution{NYU Abu Dhabi}
  \city{Abu Dhabi}
  \country{UAE}
}


\begin{abstract}
The manner in which different racial and gender groups are portrayed in news coverage plays a large role in shaping public opinion. As such, understanding how such groups are portrayed in news media is of notable societal value, and has thus been a significant endeavour in both the computer and social sciences. Yet, the literature still lacks a longitudinal study examining both the frequency of appearance of different racial and gender groups in online news articles, as well as the context in which such groups are discussed. To fill this gap, we propose two machine learning classifiers to detect the race and age of a given subject. Next, we compile a dataset of 123,337 images and 441,321 online news articles from New York Times (NYT) and Fox News (Fox), and examine representation through two computational approaches. Firstly, we examine the frequency and prominence of appearance of racial and gender groups in images embedded in news articles, revealing that racial and gender minorities are largely under-represented, and when they do appear, they are featured less prominently compared to majority groups. Furthermore, we find that NYT largely features more images of racial minority groups compared to Fox. Secondly, we examine both the frequency and context with which racial minority groups are presented in article text. This reveals the narrow scope in which certain racial groups are covered and the frequency with which different groups are presented as victims and/or perpetrators in a given conflict. Taken together, our analysis contributes to the literature by providing two novel open-source classifiers to detect race and age from images, and shedding light on the racial and gender biases in news articles from venues on opposite ends of the American political spectrum.
\end{abstract}



\maketitle

\section{Introduction}
News coverage plays a significant role in shaping public opinion through the frequency and language with which it describes those involved in a particular event~\cite{mccombs1972agenda, mccombs1993evolution, reynolds2002news}. For instance, studies throughout the 21st century have highlighted that local news disproportionately present people of color as criminals and White individuals as victims in comparison to crime reports~\cite{dixon2000overrepresentation, dixon2015changing, dixon2000race, dixon2003portrayal}.  A similar phenomenon exists regarding gender bias, with many studies evaluating both the proportion of images on news websites that include female individuals, as well as the language in which females are described within articles~\cite{d2020bits, cullity2009gender, rao2021gender, dacon2021does}. These studies largely find that females are underrepresented, marginalized, and suffer from socially-constructed biases. 

The implications of such potentially biased representations of racial and gender minorities lie in the impact that such representation could have on individuals' perceptions of such groups. Indeed, many studies have noted that exposure to racially distorted portrayals of crime, for instance, reinforces negative stereotypes of racial minorities~\cite{banjo2018us, gilens2009americans, gilliam1996crime, johnson1997race, hamborg2019automated}. Specifically, through the ample literature on cultivation theory, it is widely theorized that the consistency and frequency of the messages relayed from media sources shape both individual and societal values~\cite{shrum2017cultivation, stein2021every, hermann2023cultivation}. Given that the American consumer spends around eight hours of their day on digital media~\cite{Cramer-Flood_2023}, and that the vast majority of news is now consumed online~\cite{Pew}, it is imperative to understand both the frequency with which different groups are presented in online news coverage, as well as the manner and context in which they are presented. Rather than relying on laborious human annotation, which consequently necessitates relatively small sample sizes, recent literature has relied on advancements in image classification and natural language processing techniques to analyze such biases at scale, with text and image sample sizes growing several orders of magnitude over the last decade.

Against this background, we make two key contributions:
\begin{itemize}
    \item We introduce and validate two state-of-the-art models for classifying images of individuals into their respective racial group and age-group, outperforming available models.
    \item We utilize these models, in addition to other established text-based classifiers, to analyze the representation of racial and gender groups temporally in the online articles of two popular news outlets on opposite sides of the US political spectrum over the course of a decade. 
\end{itemize} 

To this end, we compile a dataset of 123,337 images and 441,321 online news articles from New York Times and Fox News by scraping the websites of these venues, and examine racial and gender representation both in images and text to answer the following research questions:

\begin{itemize}
    \item \textbf{RQ1:} At what frequency are different racial and gender groups portrayed in images hosted on NYT and Fox, and do these venues differ in their inclusion of these groups?
    \item \textbf{RQ2:} In what emotional tilt and textual sentiment are different racial groups portrayed?
    \item \textbf{RQ3:} How do different racial groups differ with regards to the article topics they appear most frequently in?
    \item \textbf{RQ4:} At what frequency do different racial group pairs appear as victims or perpetrators in a given article?
\end{itemize}

\textbf{Summary of main findings:}
Overall, we find that racial and gender minorities are under-represented in both NYT and Fox in terms of their frequency of appearance in images embedded in online news articles. Moreover, when such minorities do appear, they appear less prominently compared to their majority counterparts. Comparing NYT and Fox directly in this regard, we find that NYT generally features images of racial and gender minorities more frequently than Fox, with Black/African Americans in sports-related images being the sole exception.

In text, we find differences in the emotion conveyed in news articles published by both venues, with NYT publishing articles more aligned with the emotion of sadness and fear, and Fox publishing articles more in line with anger or joy. Furthermore, we find that the sentimental charge of Fox articles was more polarized relative to NYT (i.e., NYT coverage had more balanced coverage with regards to the quantity of positive and negative sentiment articles concerning a given racial group, relative to Fox). We also find that both venues' coverage of certain topics was largely homogeneous with regard to the racial group portrayed (e.g., 95.4\% of articles on terrorism in Fox were concerned with Middle Eastern individuals or groups). Lastly, focusing on articles that depict a victim and a perpetrator, we find that both venues often depict a given racial minority group, such as Asian or Middle Eastern, as the perpetrator in articles in which the group is also portrayed as the victim. However, in the case of Black and White individuals, this was not the case, with White individuals being more likely to be depicted as perpetrators against Black victims.

\vspace{-2mm}
\section{Related Work}
\textbf{Biases in media:} Understanding racial and gender biases in various forms of media has been an ongoing endeavour in the computational and social sciences. Prior work, for instance, has examines such biases in fashion~\cite{aldahoul2024inclusive, han2015images, pilane2016miss, pious1997racial, reddy2018critical, sengupta2006reading, wasylkiw2009all}, film~\cite{aldahoul2024inclusive, amaral2020long, topaz2022race, smith2003popular, smith2012gender, smith2010assessing}, and advertisements~\cite{olsson2018does, aldahoul2024inclusive}. Yet, less is understood regarding the existence of such biases in the images embedded in online news articles. Indeed, the majority of previous work on racial biases in the news has examined the portrayal of minority racial groups in television news coverage~\cite{dixon2000overrepresentation, dixon2000race, dixon2015changing}. With regards to gender bias, a handful of studies have examined gender representation in news images~\cite{jia2015measuring, kjeldsen2024broken}, finding that gender bias differs by topic and type of news venue. Yet, these studies largely examine a narrow temporal scope, focusing on articles or images published over the course of a year at most. On the other hand, analyzing gender bias in text in the context of news articles has been well studied in the literature, particularly as natural language processing techniques have evolved over the last decade~\cite{dacon2021does, lafrance2016analyzed, davis2022gender, rao2021gender, niven2016gender, d2020bits}. Nevertheless, to the best of our knowledge, there has been no longitudinal study examining racial biases in news article text and images to date.

\textbf{Image classification:} Recent advancements in classifying gender, race, and age from images have been made to address several challenges from biased datasets, varying lightning conditions, and pose variations. Most notably, Karkkainen et al.~~\cite{karkkainen2021fairface} introduced the FairFace dataset, consisting of 108,501 images balanced across seven racial groups, namely Black, East Asian, Indian, Latinx, Middle Eastern, Southeast Asian, and White. Furthermore, this dataset also included labels for the gender and age group of a given subject, allowing for subsequent work on classifying such attributes from images. With the introduction of this dataset, numerous models have been introduced to classify race, gender, and age from images~\cite{karkkainen2021fairface, he2016deep, aldahoul2024ai, cao2018vggface2, schroff2015facenet, khurana2024leveraging, kalkatawi2024ethnicity, radford2021learning}. For instance, Kalkatawi et al.~ introduce Multi-Axis Vision transformer (Max Vit) proposed by Google, a hybrid transformer-based model capable of feature extraction and classification for ethnicity recognition. Max Vit has the backbone structure of a self-attention mechanism, known as multi-axis self-attention (Max-SA),  which is capable of capturing both local and global semantic relationships between data segments~\cite{karkkainen2021fairface}. Max Vit outperformed other convolutional neural network models, including ResNet50, EfficientNet-B2, and VGG16, by achieving the highest accuracy with a smaller number of model parameters~\cite{karkkainen2021fairface}. Radford et al.~~\cite{radford2021learning} tested the performance of the CLIP model across intersectional race and gender categories defined in the FairFace dataset. In their implementation, natural language is used to leverage learned visual concepts to enable zero-shot transfer of the model to various downstream tasks. They showed competitive results compared to a fully supervised baseline without the need for any training dataset. In our study, we specifically build upon the work conducted by AlDahoul et al.\ ~\cite{aldahoul2024ai}, which utilized a VGGFace ResNet-50 convolutional neural network to extract embedding vectors from images in the FairFace dataset, and added a support vector machine classifier to classify the resulting vectors into racial, age, and gender groups.

\vspace{-0.5em}
\section{Methods}

\textbf{Data:} Our analysis of racial and gender representation in online news outlets is two-fold, first with regards to the images embedded in articles, and second with regard to the textual content of the articles. When analyzing images, we focus on four neutral categories of news articles, namely those of Food, Travel, Sport and Arts. For the first three of these categories, both websites considered in our analysis (i.e., NYT and Fox News) host designated subdomains with these specific titles. We group the two New York Times categories of Arts and Lifestyle, as well as the two Fox News categories of Entertainment and Lifestyle into a single category called Arts, since they discuss similar topics such as cinema, TV, and celebrities. We collect all of the articles listed in these categories from 2012 to 2022 and identify any images embedded therein. Focusing on the images that depict a person left us with 123,227 images in total. As for the textual analysis, in addition to the articles collected above, we also collect all of the articles under the categories of Opinion, Politics, Science, Technology, U.S., and World from the years 2012 to 2022. This was done by utilizing the Wayback Machine to access daily historical versions of each website~\cite{notess2002wayback}. In total, this amounted to 184,551 unique articles from Fox News and 256,770 unique articles from New York Times.

\textbf{Image Classification:}
To classify the race and age of a given subject from an image, we first establish a performance baseline by utilizing the FairFace dataset~\cite{karkkainen2021fairface} to compare against state-of-the-art race classification models from the literature. FairFace is one of the largest publicly-available datasets of face images. Specifically, the datasets annotates the race of the subject into one of seven categories, namely, Black, East Asian, Indian, Latinx, Middle Eastern, Southeast Asian, and White, as well as the gender of the subject as either male or female. The models that we tested were CLIP~\cite{radford2021learning}, FairFace ResNet34~\cite{karkkainen2021fairface}, FaceNet-SVM~\cite{aldahoul2024ai}, VGGFaceResnet50~\cite{aldahoul2024ai}, EfficientNet~\cite{khurana2024leveraging}, and Vision transformer~\cite{kalkatawi2024ethnicity}. All the aforementioned models other than EfficientNet and Vision Transformer were used as pre-trained models evaluated on 10,954 samples of testing data from the FairFace dataset. We also fine-tuned all layers in EfficientNet and Vision Transformer using training data from FairFace, and evaluated the resulting fine-tuned models with 10,954 samples of testing data from FairFace. The results of the aforementioned models are listed in Table~\ref{tab:race_comparision}. As a further point of comparison, we replace the top layers in each of the EfficientNet and Vision Transformers models with a support vector machine (SVM) classification step. We find that the combination of EfficientNet with an SVM classifier, as well as the combination of Vision Transformer with an SVM classifier, outperform the aforementioned models by 2\% and 1\%, respectively; see Table~\ref{tab:race_comparision}.

Next, we propose a novel solution for racial group classification, which consists of four stages:
\begin{enumerate}
\item \textbf{Face detection stage:} To detect the face of a given subject, a Multi-task Cascaded Convolutional Neural Network (MTCNN)~\cite{zhang2016joint} was selected to balance high detection accuracy and run-time speed. This MTCNN was configured to filter boundary-boxes and keep ones with a confidence score exceeding 0.9.

\item \textbf{Face embedding generation stage:} In this stage, two embedding models, namely VGGFaceResNet50
Convolutional Neural Networks (VGGFaceResNet50 CNN)~\cite{cao2018vggface2} and Vision Transformer~\cite{dosovitskiy2020image}, were used to extract informative features from the faces detected in the previous stage. 
First, we utilized a pre-trained VGGFaceResNet50 CNN that was trained on MS-Celeb-1M~\cite{guo2016ms} and VGGFace2~\cite{cao2018vggface2}. The top layers of VGGFaceResNet50 CNN were removed to extract a first embedding vector of 2048 dimensions. The images were resized to 224x224 pixels before being fed to the CNN as input. 
Subsequently, we fine-tuned all the layers of large version of Vision Transformer (VIT)~\cite{google/vit-large-patch16-224} on the FairFace dataset~\cite{karkkainen2021fairface}. The fine-tuned VIT was used after removing the top layers to extract a second embedding vector of 1024 dimensions.

\item \textbf{SVM classification stage:} After retrieving the two embedding vectors, we classified the vectors extracted from each face image using two Support Vector Machine (SVM) classifiers~\cite{hearst1998support}. Specifically, the two embedding vectors outputted from VIT and VGGFaceResNet50 are fed to two SVMs that were configured to produce prediction probabilities.

\item \textbf{Averaging model predictions:} The final class decision was calculated by averaging the prediction probabilities of both SVM classifiers described in the previous stage.
 
\end{enumerate}

\begin{figure*}
    \centering
    \includegraphics[width=\linewidth]{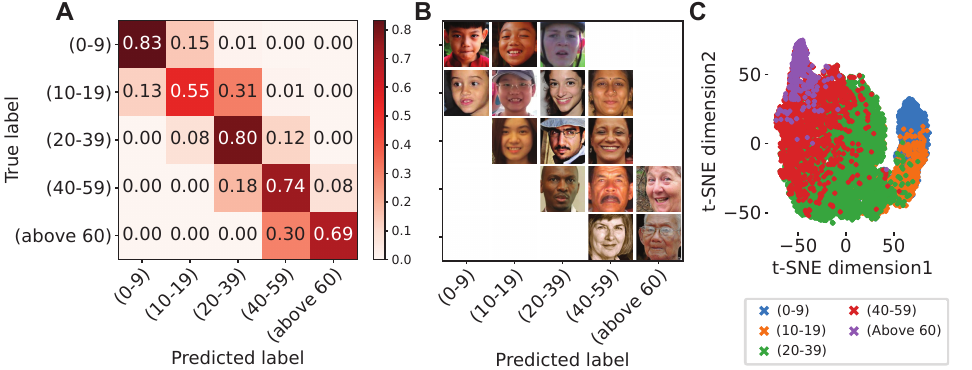}
    \caption{Summary of age group classification. (A) Confusion matrix of predictions made by the model for the different age groups in the dataset. (B) Examples of images that were correctly or incorrectly classified by the model when predicting age group. (C) Results of t-SNE dimensionality reduction of the images with regards to age group.}
    \Description{Summary of age group classification. (A) Confusion matrix of predictions made by the model for the different age groups in the dataset. (B) Examples of images that were correctly or incorrectly classified by the model when predicting age group. (C) Results of t-SNE dimensionality reduction of the FairFaice images with regards to age group.}
    \label{fig:age_classifier}
\end{figure*}

The results of our proposed racial group classification model for each racial group are shown in Table~\ref{tab:race_classification_report}. As can be seen, the model achieves an average accuracy of 75\% and an average macro F1 score of 75\%, exceeding that of previous work. A summary of the model's results can be seen in Figure~\ref{fig:race_classifier}. More specifically, Figure~\ref{fig:race_classifier}A illustrates the confusion matrix of the model for different racial group predictions, Figure~\ref{fig:race_classifier}B illustrates examples of correctly and incorrectly classified faces, while Figure~\ref{fig:race_classifier}C depicts the results of t-SNE dimensionality reduction of the racial group prediction vectors on FairFace images. For the purposes of our subsequent analysis of racial representation in news images, we combined the Southeast Asian and East Asian classified racial groups into one group named ``Asian''. The proposed model is open-sourced and can be found at the following link: \url{https://github.com/comnetsAD/NYT_vs_FoxNews}

For age group classification, we fine-tuned all the layers of the Base version of Vision Transformer (VIT)~\cite{google/vit-base-patch16-224} on the FairFace dataset~\cite{karkkainen2021fairface}. The fine-tuned VIT was used to produce the age group categories as probabilities at the output. The results of our age classification model for each age group are shown in Table~\ref{tab:age_classification_results}. As can be seen, the model achieves an average accuracy of 76\% and an average macro F1 score of 71\%. We compare our age classifier with Amazon Rekognition (AWS)~\cite{Amazon_Rekognition}, which estimates an age range for faces detected in an image. As shown in Table~\ref{tab:age_comparison}, our classifier outperforms AWS in terms of all performance measures. Figure~\ref{fig:age_classifier}A shows the confusion matrix of model predictions, Figure~\ref{fig:age_classifier}B illustrates examples of correctly and incorrectly classified faces, while Figure~\ref{fig:age_classifier}C illustrates results of t-SNE dimensionality reduction of age group prediction vectors on images in the FairFace dataset. The proposed model can be found at the following link: \url{https://github.com/comnetsAD/NYT_vs_FoxNews}

\begin{figure*}
    \centering
    \includegraphics[width=\linewidth]{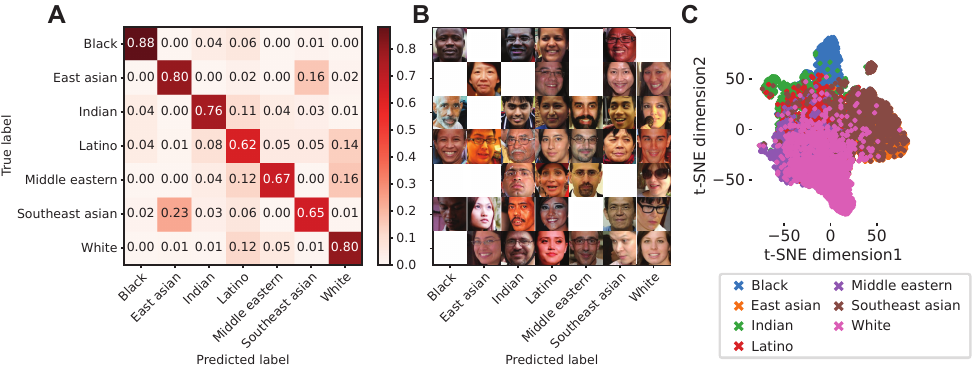}
    \caption{\textbf{Summary of racial classifier results}. (A) Confusion matrix of predictions made by the model for different racial groups. (B) Examples of images that were correctly or incorrectly classified by the model when predicting racial group. (C) Results of t-SNE dimensionality reduction of the images with regards to racial group.}
    \label{fig:race_classifier}
    \Description{(A) Confusion matrix of predictions made by the model for different racial groups. (B) Examples of images that were correctly or incorrectly classified by the model when predicting racial group. (C) Results of t-SNE dimensionality reduction of the FairFace images with regards to racial group.}
\end{figure*}

\renewcommand\tabularxcolumn[1]{>{\Centering}m{#1}}

\begin{table}[!ht]
    \centering
    \begin{tabularx}{\linewidth}{L|cccc}
        \textbf{Models} & \textbf{Accuracy} & \textbf{Precision} & \textbf{Recall} & \textbf{F1-Score} \\ \hline 
        FairFace ResNet34~\cite{karkkainen2021fairface} & 0.72 & 0.72 & 0.71 & 0.72 \\ \hline
        Zero shot CLIP~\cite{radford2021learning} & 0.64 & 0.67 & 0.65 & 0.65 \\ \hline
        Google FaceNet + SVM~\cite{aldahoul2024ai} & 0.69 & 0.69 & 0.68 & 0.68 \\ \hline
        Oxford VGGface ResNet 50~\cite{aldahoul2024ai} & 0.73 & 0.72 & 0.72 & 0.72 \\ \hline
        EfficientNetB7 with fine-tuned layers~\cite{khurana2024leveraging} & 0.70 & 0.70 & 0.70 & 0.70 \\ \hline
        Base VIT with fine-tuned layers~\cite{kalkatawi2024ethnicity} & 0.70 & 0.70 & 0.69 & 0.69 \\ \hline
        Large VIT with fine-tuned layers~\cite{kalkatawi2024ethnicity} & 0.71 & 0.72 & 0.70 & 0.71 \\ \hline
        VIT large + SVM & 0.72 & 0.72 & 0.72 & 0.72 \\ \hline
        EfficientNetB7 + SVM & 0.72 & 0.72 & 0.71 & 0.71 \\ \hline
        \textbf{VGGface ResNet + Large VIT (Proposed + Utilized)} & \textbf{0.75} & \textbf{0.75} & \textbf{0.74} & \textbf{0.74} \\ \hline
    \end{tabularx}
    \caption{A comparison between previous work and the proposed model with regards to racial group image classification.}
    \label{tab:race_comparision}
\end{table}

\begin{table}[!ht]
    \centering
    \begin{tabular}{l|cccc}
        \textbf{Race} & \textbf{Precision} & \textbf{Recall} & \textbf{F1-score} & \textbf{Support} \\ \hline 
        Black & 0.89 & 0.88 & 0.88 & 1556 \\ 
        East Asian & 0.76 & 0.80 & 0.78 & 1550 \\ 
        Indian & 0.79 & 0.76 & 0.77 & 1516 \\ 
        Latinx & 0.57 & 0.62 & 0.60 & 1623 \\ 
        M. Eastern & 0.75 & 0.67 & 0.71 & 1209 \\ 
        Southeast Asian & 0.69 & 0.65 & 0.67 & 1415 \\ 
        White & 0.78 & 0.80 & 0.79 & 2085 \\ \hline \hline
        \textbf{Macro Avg} & 0.75 & 0.74 & 0.74 & 10954 \\ 
        \textbf{Weighted Avg} & 0.75 & 0.75 & 0.75 & 10954 \\\hline \hline
        \textbf{Accuracy} & \multicolumn{4}{c}{\textbf{0.75}}  \\ \hline
    \end{tabular}
    \caption{A summary of the results for racial group image classification.}
    \label{tab:race_classification_report}
\end{table}

\begin{table}[!ht]
    \centering
    \begin{tabular}{l|cccc}
        \textbf{Age} & \textbf{Precision} & \textbf{Recall} & \textbf{F1-score} & \textbf{Support} \\ \hline
        (0-9) & 0.89 & 0.83 & 0.86 & 1553 \\ 
        (10-19) & 0.49 & 0.55 & 0.52 & 1181 \\ 
        (20-39) & 0.85 & 0.80 & 0.83 & 5630 \\ 
        (40-59) & 0.66 & 0.74 & 0.70 & 2149 \\ 
        (Above 60) & 0.64 & 0.69 & 0.67 & 439 \\ \hline \hline
        \textbf{Macro Avg} & 0.71 & 0.72 & 0.71 & 10952 \\ 
        \textbf{Weighted Avg} & 0.77 & 0.76 & 0.76 & 10952 \\ \hline \hline
        \textbf{Accuracy} & \multicolumn{4}{c}{\textbf{0.76}}  \\ \hline
    \end{tabular}
    \caption{A summary of the results for age group image classification.}
    \label{tab:age_classification_results}
\end{table}

\renewcommand\tabularxcolumn[1]{>{\Centering}m{#1}}

\begin{table}[!ht]
    \centering
    \begin{tabularx}{\linewidth}{L|cccc}
        \textbf{Models} & \textbf{Accuracy} & \textbf{Precision} & \textbf{Recall} & \textbf{F1-Score} \\ \hline 
        AWS classifier~\cite{Amazon_Rekognition} & 0.72 & 0.68 & 0.61 & 0.63 \\ \hline
        \textbf{Tuned Base Vision Transformer} &\textbf{ 0.76} & \textbf{0.71} & \textbf{0.72} & \textbf{0.71} \\ \hline
    \end{tabularx}
    \caption{A comparison between previous work (AWS) and the proposed model for age group image classification.}
    \label{tab:age_comparison}
\end{table}

Lastly, to classify gender from images, we utilize the model proposed by AlDahoul et al.~\cite{aldahoul2024ai}. A summary of the model's performance can be found in Figure~\ref{fig:gender_classifier} in the Appendix.

\textbf{Text classification:} In our analysis of online news articles, we make use of several machine learning models to classify text across a number of dimensions. First, to classify the emotional category of a body of text, we use a fine-tuned version of the Emotion DistilRoBERTA-base model~\cite{hartmann2022emotionenglish}. This model classifies text into one of the following seven emotion categories: Neutral, Disgust, Fear, Joy, Anger, Sadness, and Surprise. This model was selected as it has been shown to outperform other transformer models for the purpose of recognizing emotion from text~\cite{adoma2020comparative}. As for classifying sentiment, we use SiEBERT~\cite{hartmann2023more}, which is a fine-tuned checkpoint of RoBERTa-large~\cite{robertalarge}, classifying text as either of positive or negative sentiment, and is considered to be the state-of-the-art for classifying sentiment from text.

To classify both text category as well as racial group discussed in a body of text, we use the Bart-large model~\cite{bartlarge}. This model takes in a body of text, as well as a set of categories which the text can be classified under. For classifying text category, we provided the following list of 25 categories: Animals, Agriculture, Celebrations, Disaster, Disease, Economics, Education, Entertainment, Environment, Finance, Food, Health, Immigration, Inventions, Manufacturing, Movie, Politics, Poverty, Science, Sport, Technology, Terrorism, Violence, War, and Weather. As for classifying racial group, we provide the six racial groups analyzed throughout our study, namely: Asian, Black, Indian, Latinx, Middle Eastern, and White. Given that gender representation in news article text is well-covered in the literature~\cite{dacon2021does, lafrance2016analyzed, davis2022gender, rao2021gender, niven2016gender, d2020bits}, we focus our textual analysis on the representation and portrayal of different racial groups. 

Lastly, to gain insight into the portrayals of a given racial group in a given article, we identify whether the article is constructed in a manner such that it includes a victim and a perpetrator, and if so, what racial group falls on either end of this relationship. To this end, we use GPT-4 with the prompt described in Appendix Figure~\ref{fig:gpt4ojudge_prompt}. Examples of classification model outputs for a sample of articles is shown in Appendix Figure~\ref{fig:text_samples}.

\textbf{Model validation:} 
The outputs of both the image and text-based classification models were validated by five independent coders, the results of which can be found in Table~\ref{table:human_annotation}. In the case of image classification, we verify the accuracy of the model by annotating a sample of 200 images per group and find that the accuracy of the model, computed as the proportion of images in which the majority vote of the coders matches the model's output, is at least 70\% for racial, age, and gender group. Furthermore, Cohen's $\kappa$ was computed between the rater majority vote and the model's prediction to measure rater-model agreement for each of these classification tasks, achieving a score $\geq$ 0.65 in all tasks, which indicates moderate agreement. In each case, inter-rater agreement was also computed using Krippendorf's $\alpha$, with all classification tasks receiving an $\alpha \geq 0.62$, which indicates strong agreement. 

For text classification, we follow the same approach outlined above, verifying the results with five independent annotators with 200 text samples per output label for each classification task. As can be seen in Table~\ref{table:human_annotation}, all text classification tasks achieved an accuracy $\geq$ 0.71 and an F1-Score  $\geq$ 0.72. Similarly, all tasks received a Cohen's $\kappa \geq 0.66$ indicating strong rater-model agreement. Lastly, we test inter-rater reliability and find that all tasks had a Krippendorf's $\alpha \geq 0.56$, indicating moderate to strong agreement. 

\begin{table}[]
\begin{tabular}{l|l|cccc}
& \multicolumn{1}{c}{} & \textbf{\begin{tabular}[c]{@{}c@{}}Kripp's\\ $\alpha$\end{tabular}} & \textbf{F1} & \textbf{\begin{tabular}[c]{@{}c@{}}Cohen's\\ $\kappa$\end{tabular}} & \textbf{Acc.} \\ \hline
\multirow{6}{*}{\textbf{Text}} & \textbf{Victim}  & 0.74 & 0.72& 0.66 & 0.71\\
& \textbf{Perpetrator} & 0.73& 0.83& 0.75& 0.79\\
& \textbf{Emotion}& 0.88& 0.98& 0.97& 0.98\\
& \textbf{Category} & 0.56& 0.85& 0.85& 0.85\\
& \textbf{Sentiment}& 0.73& 0.88& 0.75& 0.88\\
& \textbf{Race} & 0.83& 0.72& 0.70& 0.73\\ \hline
\multirow{3}{*}{\textbf{Images}} & \textbf{Gender} & 0.90& 0.98& 0.96& 0.98\\
& \textbf{Age} & 0.62& 0.86& 0.83& 0.86\\
& \textbf{Race} & 0.62& 0.73& 0.65& 0.70   \\ \hline
\end{tabular}
\caption{Human annotation results for the different text and image models utilized in this study. In the first column, Krippendorf's $\alpha$ is computed to measure inter-rater reliability between the five independent coders. In the second, third, and fourth columns, F1-Score, Cohen's $\kappa$, and Accuracy are computed between the rater majority vote and the respective model's prediction, respectively.}
\vspace{-8mm}
\label{table:human_annotation}
\end{table}

\section{Results}

\begin{figure}[htbp!]
    \centering
    \includegraphics[width=\linewidth]{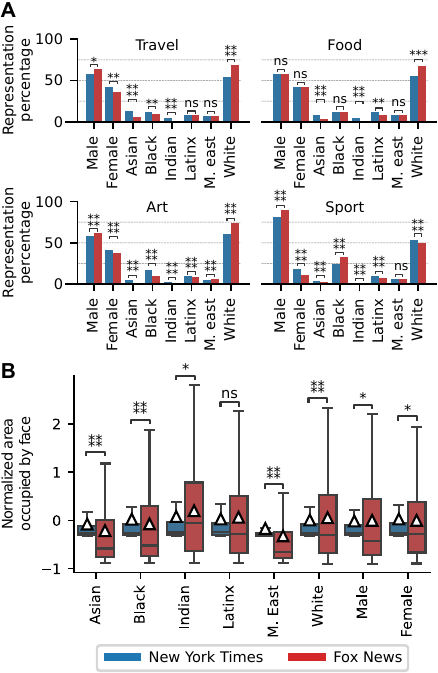}
    \caption{Race and gender in news images. (A) Racial and gender representation in the photos included in articles under the Art, Sport, Food, and Travel categories. Chi-squared tests are used to measure differences between NYT and Fox. (B) Normalized area occupied by face of different racial and gender groups across all the aforementioned categories. Independent t-tests are conducted on normalized area values of NYT and Fox. $ns: p > 0.05$, $*: p < 0.05$, $**: p < 0.01$, $***: p < 0.001$, $****: p < 0.0001$}
    \Description{Race and gender in news images. (A) Racial and gender representation in the photos included in articles under the Art, Sport, Food, and Travel categories. Chi-squared tests conducted to measure differences between NYT and Fox. (B) Normalized area occupied by face of different racial and gender groups across all the aforementioned categories. Independent t-tests conducted on normalized area values of NYT and Fox.}
    \vspace{-7mm}
    \label{fig:representation}
\end{figure}

\textbf{RQ1:} We start by analyzing the 123,227 images embedded within the articles from four different categories, namely those in the Art, Sport, Food, and Travel categories. Each of these categories are listed on the websites of both New York Times (NYT) and Fox News (Fox). Figure~\ref{fig:representation}A depicts the percentage of images falling under each gender and racial group for each news outlet. Beginning with gender, as can be seen in all four categories, males were more represented on average than females for both news outlets. This difference is most stark in the Sports category, with males accounting for 89\% of images in Fox articles and 81.5\% of images in NYT. Comparing NYT and Fox directly, we see that Fox more frequently featured images of men in all categories except Food, where there was no statistically significant difference between the two outlets.

With regards to race, images including White individuals accounted for the vast majority of images in all categories. However, while the representation of racial minorities were roughly balanced in the Art, Food, and Travel categories, Black individuals represented a larger proportion of images in the Sport category, accounting for 33\% and 25\% of images in Fox and NYT, respectively. Comparing NYT and Fox directly, we see that NYT featured more images of racial minority groups (non-white individuals) more frequently than Fox. Indeed, of the 20 instances in which racial minority representation is compared, NYT more frequently portrayed racial minorities in 13 such instances, compared to one such instance for Fox (Black representation in the Sports category). In contrast, in three of the four article categories, Fox included significantly more images of White individuals compared to NYT, with the one exception being the Sports category. Chi-squared test values to compare representation rates between the two venues can be found in Table~\ref{table:image_repr_sig} in the Appendix section. An analysis of the representation across age groups utilizing the model proposed in this study can be found in Appendix section~\ref{sec:age}.

Yet, mere appearance in an image embedded in an article may obscure the prominence with which the subject is portrayed. To address this aspect of representation, we also analyze the size of the image in which a given subject is portrayed.  Here, we first compute the area as the product of the height and width of a given image in pixels. We then compute the normalized area occupied by a given image as the z-score of the image's area across all images published by the image's respective venue. This is to control for any variations in image sizes across the two news outlets. The results of this analysis are depicted in Figure~\ref{fig:representation}B. As can be seen, when racial minorities do appear in an image, their appearance is often more prominent on average in NYT compared to Fox. This was the case for three of the five racial minority groups analyzed, namely, for Asian, Black and Middle Eastern individuals. In contrast, Indians appeared significantly more prominently in Fox, and there were no statistically significant differences in the prominence of Latinx individuals across the two news outlets. For gender, we find that males are portrayed more prominently in Fox, while females are more prominent in NYT. 

\begin{figure}[htbp!]
    \centering
    \includegraphics[width=\linewidth]{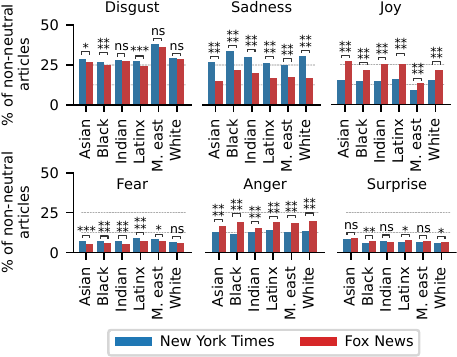}
    \caption{Proportion of non-neutral articles which mention a given race and convey a particular emotion. Chi-squared tests are used to compare the proportion of articles with a given emotional tilt in NYT and Fox News. $ns: p > 0.05$, $*: p < 0.05$, $**: p < 0.01$, $***: p < 0.001$, $****: p < 0.0001$}
    \Description{Proportion of non-neutral articles which mention a given race and convey a particular emotion. Chi-squared tests conducted between proportion of articles with a given emotional tilt in NYT and Fox.}
    \label{fig:emotion}
\end{figure}

\textbf{RQ2:} Having analyzed the images embedded in these articles, we next turn to the textual content of articles published by these venues. Here, we expand our analysis to focus on all articles published by these venues in the decade between the years 2012 and 2022 in the four aforementioned categories, as well as those in the Opinion, Politics, Technology, U.S., and World categories. First, we apply emotion classification on all articles which included a reference to an individual or group of a particular race (see Methods for more details). We find that the majority of news articles published by both venues primarily applied a neutral tone (see Supplementary Table~\ref{table:emotion_counts} for the number of articles falling under each emotion class). Figure~\ref{fig:emotion} focuses on articles which did not fall under the ``neutral'' category. As can be seen, articles published by Fox exhibited more anger, joy, and surprise amongst all racial groups. On the other hand, articles in NYT included language more associated with fear and sadness. These results largely match similar previous work analyzing emotional affect in news articles~\cite{rozado2022longitudinal}. Across racial classes, language associated with disgust was most associated with Middle Easterners, with no statistically significant differences between the two venues in this regard. Indeed, disgust was the most prevalent non-neutral emotional tone exhibited by both venues; see Table~\ref{table:emotion_sig} for statistical test results in the Appendix.

\begin{figure}[htbp!]
    \centering
    \includegraphics[width=\linewidth]{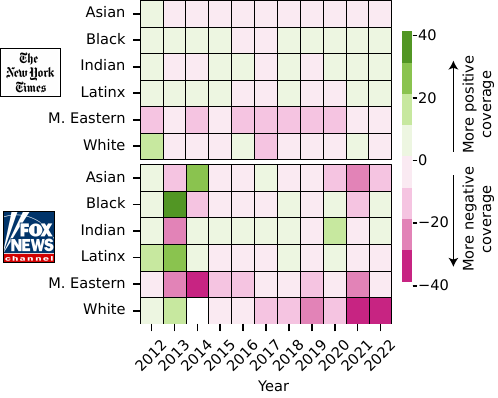}
    \caption{Normalized sentiment difference of articles which mention a particular racial group in New York Times and Fox News. Darker shades of green represent a more positive average sentiment, while darker shades of pink represent a more negative average sentiment.}
    \Description{Normalized sentiment difference of articles which mention a particular racial group in New York Times and Fox News. Dark shades of green represent a more positive average sentiment, while dark shades of pink represent a more negative average sentiment.}
    \vspace{-3mm}
    \label{fig:sentiment}
\end{figure}

We also investigate the temporal differences in the proportion of the articles which represent a given racial group positively and those that represent them negatively in both news outlets, the results of which can be seen in Figure~\ref{fig:sentiment}. Here, cells with colors closer to white denote years with more balanced coverage of a particular racial group (i.e, the difference between the proportion of positive and negative coverage is close to 0), while those with a stronger hue denote either a more positive (green) or negative (pink) coverage. As can be seen, across all racial groups, there is a small negative bias on average across both news outlets, a phenomenon consistent with prior studies on 'negativity bias' in news coverage as a whole~\cite{rozado2022longitudinal}. Yet, comparing the two outlets reveals more polarized coverage from Fox relative to NYT, where the average absolute difference in sentiment across all racial groups in Fox was 10.36\% as opposed to 4.25\% in NYT.

\begin{figure}[htbp!]
    \centering
    \includegraphics[width=\linewidth]{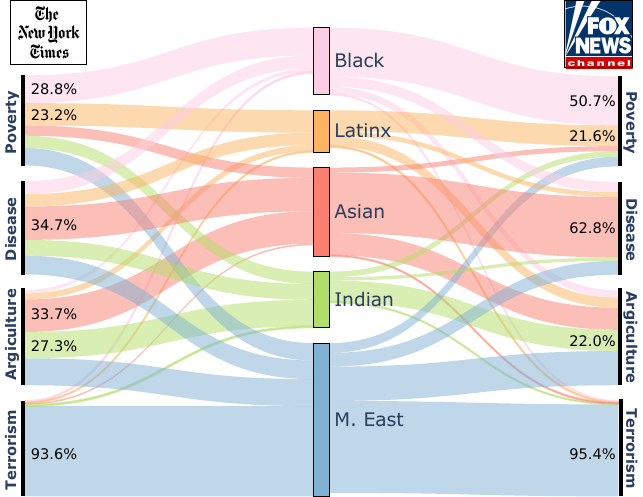}
    \caption{Sankey plot of the categories in which each racial minority group is most highly represented (e.g., 95.4\% of articles on ``Terrorism'' in Fox News concern Middle Eastern individuals or groups).}
    \Description{Sankey plot of the categories in which each racial minority group is most highly represented (e.g., 95.4\% of articles on ``Terrorism'' in Fox News concern Middle Eastern individuals or groups).}
    \label{fig:topics}
\end{figure}

\textbf{RQ3:} Next, we apply topic classification to identify the overarching topic associated with each article in the dataset. By doing so, we can identify the topics most associated with each racial group. The results of this analysis are illustrated in Figure~\ref{fig:topics}. Here, we highlight the topic most heavily associated with each of the five non-White racial groups. As can be seen, the topic-race pair with the strongest association across articles from both venues was Terrorism and the Middle Eastern racial group, as 93.6\% and 95.4\% of articles on Terrorism were associated with this group in NYT and Fox, respectively. Disease was the topic most associated with the Asian racial group, with more than 60\% of articles on the topic being associated with Asians in Fox compared to 34.7\% of articles in NYT. As for Black individuals, poverty was the most highly associated topic, with 50.7\% of articles in Fox on the topic being associated with African Americans. Finally, agriculture was most highly associated with the Latinx and Indian communities in both venues. The proportion of articles falling under each category per racial group for NYT and Fox can be seen in Supplementary Tables~\ref{table:nyt_category_counts} and ~\ref{table:fox_category_counts}, respectively. 

\begin{figure}[htbp!]
    \centering
    \includegraphics[width=\linewidth]{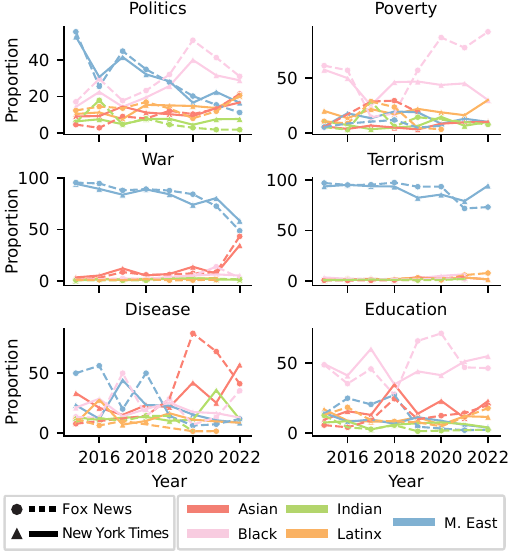}
    \caption{Temporal analysis of topics most associated with racial minority groups. For each topic shown, the proportion of articles associated with a given racial group in New York Times (triangles and solid lines) and Fox News (circles and dashed lines).}
    \Description{Temporal analysis of topics most associated with racial minority groups. For each topic shown, the proportion of articles associated with a given racial group in New York Times (triangle and solid lines) and Fox News (circle and dashed lines).}
    \vspace{-5mm}
    \label{fig:temporal_topics}
\end{figure}

We also examine the associations between a given topic and different racial minority groups over time. As can be seen in Figure~\ref{fig:temporal_topics}, some clear patterns emerge. For instance, in the topics of War and Terrorism, Middle Eastern individuals or groups were the most common subject throughout the decade covered in our dataset. However, in other cases, we see that the association between a given topic and a racial group was ephemeral, such as the association between Asians and Disease coinciding with the COVID-19 pandemic. However, across the majority of largely negative topics, such as War, Terrorism, Poverty, or Disease, we find that Fox News covered such topics with a given racial minority group as the subject more frequently than New York Times. For instance, in the case of the topic of Poverty, over 80\% of the articles on the topic were concerned with Black individuals or groups in Fox News, compared to roughly 40\% in New York Times between the years of 2020 and 2022.

\begin{figure}[htbp!]
    \centering
    \includegraphics[width=\linewidth]{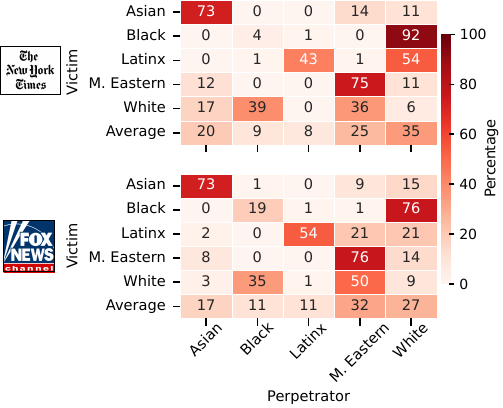}
    \caption{Out of all articles that include a victim and perpetrator, the distribution of victim-perpetrator racial pairs in New York Times and Fox News.}
    \Description{Of articles which included a victim and perpetrator, the distribution of victim-perpetrator racial pairs in New York Times and Fox News.}
    \label{fig:victim}
\end{figure}

\textbf{RQ4:} Yet, association with a particular topic such as ``War'', for instance, may obscure whether a given racial group is portrayed as the victim or the perpetrator. To gain insight into the portrayals of a given racial group in a given article, we identify whether an article is constructed in a manner where it includes a victim and a perpetrator, and if so, who falls on either end of this relationship (see Methods for more details). The results of this analysis for NYT and Fox articles can be seen in Figure~\ref{fig:victim}. Here, racial groups on the y-axis correspond to the victim in a given victim-perpetrator pair, while those on the x-axis correspond to the perpetrator. The values in the heat-map reflect the proportion of articles with the corresponding victim-perpetrator pair of all articles that include a given race as a victim. As an example, of all NYT articles in which an Asian individual is portrayed as a victim, 73\% of such articles portray Asian individuals as the perpetrator (e.g., tensions between China and Taiwan or Pakistan and India, the Myanmar Civil War). Indeed, we can see that intra-race victim-perpetrator pairs were most common in the Asian and Middle Eastern racial groups across both news outlets, with this also being the case in the Latinx group in Fox. However, for Black and White individuals, this was far from the case. For instance, in the case of Black victims, the intra-race pair only amounted to 4\% and 19\% of such articles, whereas White individuals were depicted as the perpetrators 92\% and 76\% of the time in NYT and Fox, respectively (e.g., articles discussing protests due to the killing of George Floyd in 2020, the Charleston church shootings in 2015, or broader historical recollections of slavery in the United States). Inversely, the perpetrators in articles with White victims were distributed across multiple racial groups in NYT articles. In Fox, however, Middle Eastern individuals or groups were portrayed as the perpetrators in 50\% of such articles.  

\section{Discussion and Future Work}
With the consumption of digital media in general, and online news in particular, rising considerably over the last decade~\cite{Pew}, it is crucial to understand the manner in which individuals from different races, ages, and genders are portrayed in the news. This is especially of concern given the ample literature highlighting the impact of biased racial portrayals in reinforcing negative stereotypes of racial minorities~\cite{banjo2018us, gilens2009americans, gilliam1996crime, johnson1997race, hamborg2019automated}. In this study, we first develop novel image classifiers to detect the race and age of a given subject, and subsequently utilize these classifier, in addition to other state-of-the-art models from recent literature, to paint a detailed picture of the state of racial and gender representation in two mainstream news outlets, namely New York Times and Fox News. We find that, across both news outlets, racial minorities and females are underrepresented in images, and feature less prominently with regards to the area their images occupy compared to White and Male subjects, respectively. In addition, we find statistically significant differences between the two news outlets with regards to the emotional tilt with which articles are written, as well as the sentiment polarity of these articles. 

Our work is not free from some limitations. Firstly, with regards to racial image classification, our analysis is limited to the seven racial groups analyzed, which we reduce to six groups to simplify our analysis. This is as FairFace only includes these seven groups as race annotation labels in their dataset. Likewise, gender was only classified into male or female due to the same reason. As such, future work may examine solutions towards building datasets of facial images with a more comprehensive set of racial and gender group labels, allowing for more fine-grained explorations of biases in representation. 

Future work building on this study may further examine the multitude of other news outlets falling further along each side of the political spectrum. While previous work~\cite{rozado2022longitudinal} has comprehensively examined the emotions exhibited in the language of news articles from hundreds of venues across the political aisle, future work may further examine the interplay between such emotional tilt and the subject of the article, or the topic in which they are discussed. Furthermore, we invite the research community to utilize the models proposed in this study to examine racial representation in other contexts, such as studying the impact of racially diverse content on social media or analyzing the impact of diversity, equity, and inclusion initiatives in academic institutions.

\bibliographystyle{acm}
\bibliography{sample-base}

\section{Appendix}
\begin{figure}[!h]
    \centering
    \includegraphics[width=\linewidth]{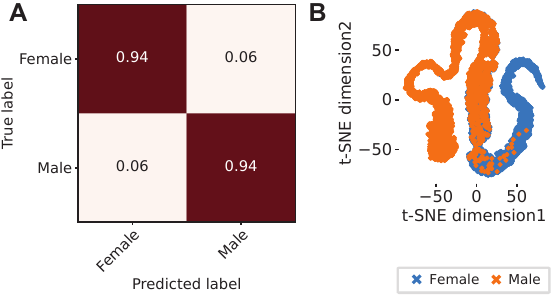}
    \caption{\textbf{Summary of gender classifier results}. (A) Confusion matrix of predictions made by the model for different racial groups. (B) Results of t-SNE dimensionality reduction of the images with regards to gender.}
    \Description{(A) Confusion matrix of predictions made by the model for different racial groups. (B) Results of t-SNE dimensionality reduction of the images with regards to gender.}
    \label{fig:gender_classifier}
\end{figure}

\begin{figure}[!bp]
    \centering
    {\small
    \begin{tcolorbox}[colback=orange!5!white, colframe=orange!75!black, title=GPT-4o Classifier Prompt, rounded corners, boxrule=1pt, boxsep=1pt]
    \textbf{[System]}
    Using the following article, answer each of the following questions using only one word.    
\newline
\newline
    \textbf{ Q1)} Does the article contain a victim? If so, what racial group does the victim belong to? Pick one of the following: ['Asian', 'Middle Eastern', 'Black', 'White', 'Indian', 'Latinx', 'Unspecified']. 
     
         If the article does not contain a victim, answer with 'No victim'.
\newline
\newline
     \textbf{Q2)} Does the article contain a perpetrator? If so, what racial group does the perpetrator belong to? Pick one of the following: ['Asian', 'Middle Eastern', 'Black', 'White', 'Indian', 'Latinx', 'Unspecified'].
     
          If the article does not contain a perpetrator, answer with 'No perpetrator'.
\newline
\newline
    Do not add any additional information from your end, answer each of the questions using only one word. 
    \newline
\newline
\textbf{    Return your response in the following JSON format:}

    \{
        'victim' : [Your response],
        
        'perpetrator' : [Your response]   
    \}
\newline
\newline
    \textbf{ARTICLE}
    
    \textbf{.........}
    \end{tcolorbox}
    }
    \vspace{-5pt}
    \caption{Prompt to determine the victim and perpetrator in a given body of text.}
    \Description{Prompt to determine the victim and perpetrator in a given body of text.}
    \label{fig:gpt4ojudge_prompt}
\end{figure}

\begin{figure}[!htbp]
    \centering
    \includegraphics[width=\linewidth]{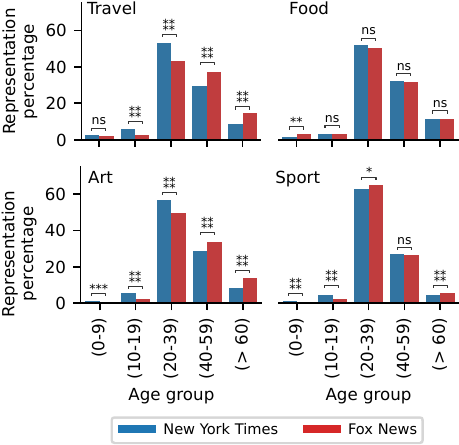}
    \caption{Age group representation in the photos included in articles under the Art, Sport, Food, and Travel categories. Chi-squared tests conducted to measure differences between NYT and Fox. $ns: p > 0.05$, $*: p < 0.05$, $**: p < 0.01$, $***: p < 0.001$, $****: p < 0.0001$}
    \Description{Age group representation in the photos included in articles under the Art, Sport, Food, and Travel categories. Chi-squared tests conducted to measure differences between NYT and Fox.}
    \label{fig:age_representation}
\end{figure}

\subsection{Age representation in NYT and Fox News}
\label{sec:age}

Having analyzed race and gender representation in NYT and Fox in RQ1, here, we look to understand the distribution of subject ages in the images embedded in the articles of these venues. To do so, we utilize the model proposed in this study, which classifies the age of a given individual into one of five age groups, namely 0-9, 10-19, 20-39, 40-59, and greater than 60 years of age (see details in the Methods section). As done in our analysis of RQ1, we focus on the images embedded in articles in the Travel, Food, Art, and Sports sections of the two news outlets.

The results of this analysis can be seen in Figure~\ref{fig:age_representation}. As shown, in the case of both venues, the 20-39 age bracket was the most represented group, followed by the 40-59 bracket. Perhaps unsurprisingly, the Sports category hosted the largest number of images in the 20-39 age bracket and the lowest number of images in the greater than 60 bracket for both venues. Comparing the two venues directly, we can see that Fox tends to skew slightly older with regards to the individuals shown in images. Taking the mean age of each bracket (70 is chosen for the greater than 60 bracket), we find that the mean age of individuals shown in NYT images is 36.83, compared to 38.74 in Fox (\textit{N} = 123,219, $p < 0.0001$).

\subsection{Supplementary Tables}

\begin{table*}[]
\begin{tabular}{l|c|c|c|c|c|c|c|c}
\textbf{Topic} & \textbf{Asian} & \textbf{Black} & \textbf{Indian} & \textbf{Latinx} & \textbf{M. East} & \textbf{White} & \textbf{Female} & \textbf{Male} \\ \hline
\textbf{Art}    & $564.93^{****}$    & $497.9^{****}$& $178.65^{****}$& $27.01^{****}$ & $25.53^{****}$  & $224.66^{****}$    & $25.22^{****}$ & $14.53^{***}$\\
\textbf{Food}   & $51.75^{****}$& $0.04$ & $28.13^{****}$ & $9.99^{**}$    & $0.0$    & $13.97^{***}$ & $0.0$  & $0.0$ \\
\textbf{Sport}  & $139.63^{****}$    & $249.7^{****}$& $21.3^{****}$  & $77.12^{****}$ & $0.63$   & $28.11^{****}$& $509.17^{****}$& $51.89^{****}$    \\
\textbf{Travel} & $84.02^{****}$& $9.49^{**}$   & $47.94^{****}$ & $0.6$   & $0.02$   & $27.36^{****}$& $8.57^{**}$    & $4.74^{*}$  \\ \hline
\end{tabular}
\caption{Chi-squared test results comparing New York Times and Fox News with regards to the proportion of images featuring a given racial or gender group in different article topics.}
\label{table:image_repr_sig}
\end{table*}

\renewcommand\tabularxcolumn[1]{>{\Centering}m{#1}}

\begin{table*}[]
\begin{tabular}{l|c|c|c|c|c|c|c}
\textbf{Venue} & \textbf{Anger} & \textbf{Disgust} & \textbf{Fear} & \textbf{Joy} & \textbf{Neutral} & \textbf{Sadness} & \textbf{Surprise} \\ \hline
\textbf{Fox News}  & 3627           & 5620             & 1248          & 4460         & 518308           & 3611             & 1435\\
\textbf{\begin{tabular}[c]{@{}l@{}}New York\\ Times\end{tabular}} & 2657           & 6094             & 1568          & 2832         & 626624           & 5888             & 1329 \\ \hline
\end{tabular}
\caption{The number of articles with a given emotional tilt in Fox News and New York Times.}
\label{table:emotion_counts}
\end{table*}

\begin{table*}[]
\begin{tabular}{l|c|c|c|c|c|c}
\textbf{Emotion}  & \textbf{Asian} & \textbf{Black} & \textbf{Indian} & \textbf{Latinx} & \textbf{M. East} & \textbf{White} \\ \hline
\textbf{Anger}    & $28.18^{****}$& $173.49^{****}$    & $27.76^{****}$ & $48.87^{****}$ & $71.84^{****}$  & $119.45^{****}$    \\
\textbf{Disgust}  & $4.27^{*}$    & $4.84^{*}$    & $1.76$  & $11.47^{***}$  & $1.61$   & $0.26$ \\
\textbf{Fear}& $14.25^{***}$ & $18.82^{****}$& $33.45^{****}$ & $30.93^{****}$ & $4.74^{*}$ & $1.21$    \\
\textbf{Joy} & $219.84^{****}$    & $126.98^{****}$    & $293.54^{****}$& $164.98^{****}$& $49.1^{****}$   & $104.68^{****}$    \\
\textbf{Sadness}  & $238.63^{****}$    & $232.36^{****}$    & $206.06^{****}$& $169.32^{****}$& $90.86^{****}$  & $314.57^{****}$    \\
\textbf{Surprise} & $1.2$ & $9.8^{**}$    & $0.77$  & $5.27^{*}$& $0.91$   & $4.42^{*}$   \\ \hline
\end{tabular}
\caption{Chi-squared test results comparing New York Times and Fox news with regards to the number of articles of a given emotional tilt with a given racial group as the subject of the article.}
\label{table:emotion_sig}
\end{table*}

\begin{table*}[!ht]
    \small
    \centering
    \begin{tabular}{c|cccccc}
    \hline
        \textbf{Category} & \textbf{Asian} & \textbf{Black} & \textbf{Indian} & \textbf{Latinx} & \textbf{M. East} & \textbf{White} \\ \hline
        \textbf{Animals} & 1.2 & 0.91 & 2.65 & 1.29 & 0.28 & 1.19 \\ 
        \textbf{Agriculture} & 2.33 & 0.19 & 3.52 & 0.31 & 0.47 & 0.16 \\ 
        \textbf{Celebration} & 0.63 & 0.18 & 1.0 & 0.3 & 0.4 & 0.63 \\ 
        \textbf{Disaster} & 4.9 & 5.16 & 6.36 & 5.84 & 5.59 & 4.15 \\ 
        \textbf{Disease} & 4.24 & 2.58 & 4.79 & 2.43 & 0.76 & 1.04 \\ 
        \textbf{Economics} & 3.09 & 2.01 & 2.02 & 2.34 & 0.56 & 2.87 \\ 
        \textbf{Education} & 2.89 & 6.86 & 2.13 & 3.02 & 0.56 & 3.99 \\ 
        \textbf{Entertainment} & 1.34 & 0.99 & 1.6 & 1.55 & 0.51 & 0.86 \\ 
        \textbf{Environment} & 4.48 & 1.49 & 3.94 & 1.69 & 0.63 & 1.37 \\ 
        \textbf{Finance} & 2.7 & 2.03 & 2.83 & 2.66 & 1.11 & 0.88 \\ 
        \textbf{Food} & 2.38 & 0.96 & 2.0 & 3.29 & 0.91 & 1.31 \\ 
        \textbf{Health} & 5.05 & 4.96 & 7.8 & 5.35 & 1.64 & 2.97 \\ 
        \textbf{Immigration} & 15.38 & 6.73 & 9.44 & 13.09 & 9.58 & 3.66 \\ 
        \textbf{Inventions} & 0.8 & 0.23 & 0.99 & 0.24 & 0.17 & 0.33 \\ 
        \textbf{Manufacturing} & 3.3 & 1.95 & 3.55 & 1.23 & 1.6 & 3.52 \\ 
        \textbf{Movie} & 0.92 & 0.88 & 0.74 & 0.46 & 0.18 & 0.72 \\ 
        \textbf{Politics} & 18.75 & 36.2 & 22.68 & 33.76 & 20.99 & 51.88 \\ 
        \textbf{Poverty} & 0.31 & 2.04 & 1.06 & 1.68 & 0.22 & 0.96 \\ 
        \textbf{Science} & 1.35 & 1.07 & 1.05 & 0.39 & 0.15 & 1.43 \\ 
        \textbf{Sport} & 2.22 & 0.88 & 2.11 & 2.0 & 0.79 & 0.98 \\ 
        \textbf{Technology} & 6.67 & 1.38 & 3.71 & 1.88 & 1.79 & 1.64 \\ 
        \textbf{Terrorism} & 0.24 & 0.54 & 0.41 & 0.29 & 6.7 & 0.8 \\ 
        \textbf{Violence} & 10.26 & 18.15 & 11.29 & 12.84 & 25.65 & 11.36 \\ 
        \textbf{War} & 3.99 & 1.48 & 1.33 & 1.74 & 18.66 & 1.11 \\ 
        \textbf{Weather} & 0.57 & 0.16 & 1.0 & 0.36 & 0.11 & 0.18 \\ \hline
    \end{tabular}
    \caption{The proportion of articles in the New York Times falling under each article category per racial group.}
    \label{table:nyt_category_counts}
\end{table*}

\begin{table*}[!ht]
\small
    \centering
    \begin{tabular}{c|cccccc}
    \hline
        \textbf{Category} & \textbf{Asian} & \textbf{Black} & \textbf{Indian} & \textbf{Latinx} & \textbf{M. East} & \textbf{White} \\ \hline
        \textbf{Animals} & 0.36 & 0.48 & 0.65 & 0.21 & 0.21 & 0.49 \\ 
        \textbf{Agriculture} & 0.69 & 0.05 & 0.96 & 0.17 & 0.16 & 0.00 \\ 
        \textbf{Celebration} & 0.36 & 0.32 & 0.86 & 0.21 & 0.42 & 0.31 \\ 
        \textbf{Disaster} & 9.79 & 8.55 & 4.33 & 5.61 & 8.71 & 6.73 \\ 
        \textbf{Disease} & 7.58 & 0.88 & 0.81 & 0.38 & 0.31 & 0.94 \\ 
        \textbf{Economics} & 6.36 & 2.66 & 2.57 & 2.81 & 0.7 & 1.78 \\ 
        \textbf{Education} & 2.45 & 6.77 & 2.72 & 2.94 & 0.41 & 4.68 \\ 
        \textbf{Entertainment} & 0.69 & 0.88 & 0.86 & 1.09 & 0.4 & 1.6 \\ 
        \textbf{Environment} & 2.74 & 0.86 & 1.51 & 0.94 & 0.36 & 0.89 \\ 
        \textbf{Finance} & 2.83 & 1.55 & 4.23 & 3.15 & 1.04 & 1.34 \\ 
        \textbf{Food} & 0.94 & 0.86 & 0.76 & 1.57 & 0.25 & 1.38 \\ 
        \textbf{Health} & 5.11 & 2.18 & 5.14 & 2.69 & 0.46 & 3.43 \\ 
        \textbf{Immigration} & 13.0 & 3.66 & 8.31 & 6.55 & 8.14 & 3.07 \\ 
        \textbf{Inventions} & 0.19 & 0.27 & 0.05 & 0.08 & 0.13 & 0.31 \\ 
        \textbf{Manufacturing} & 2.93 & 2.73 & 3.17 & 1.83 & 2.21 & 4.9 \\ 
        \textbf{Movie} & 0.65 & 0.79 & 1.01 & 0.27 & 0.13 & 1.16 \\ 
        \textbf{Politics} & 19.78 & 46.31 & 50.53 & 56.29 & 21.99 & 46.88 \\ 
        \textbf{Poverty} & 0.38 & 2.12 & 0.71 & 0.73 & 0.07 & 1.43 \\ 
        \textbf{Science} & 0.98 & 0.55 & 0.3 & 0.57 & 0.08 & 0.53 \\ 
        \textbf{Sport} & 1.19 & 0.66 & 2.07 & 1.34 & 0.32 & 1.07 \\ 
        \textbf{Technology} & 7.2 & 1.05 & 1.16 & 1.28 & 1.45 & 1.07 \\ 
        \textbf{Terrorism} & 0.29 & 0.34 & 0.96 & 0.73 & 10.64 & 1.29 \\ 
        \textbf{Violence} & 10.11 & 14.4 & 5.04 & 7.35 & 25.64 & 13.41 \\ 
        \textbf{War} & 3.2 & 1.04 & 1.16 & 1.07 & 15.73 & 1.25 \\ 
        \textbf{Weather} & 0.21 & 0.05 & 0.15 & 0.13 & 0.05 & 0.04 \\ \hline
    \end{tabular}
    \caption{The proportion of articles in Fox News falling under each article category per racial group.}
    \label{table:fox_category_counts}
\end{table*}

\begin{figure*}[!htbp]
    \centering
    \includegraphics[width=\linewidth]{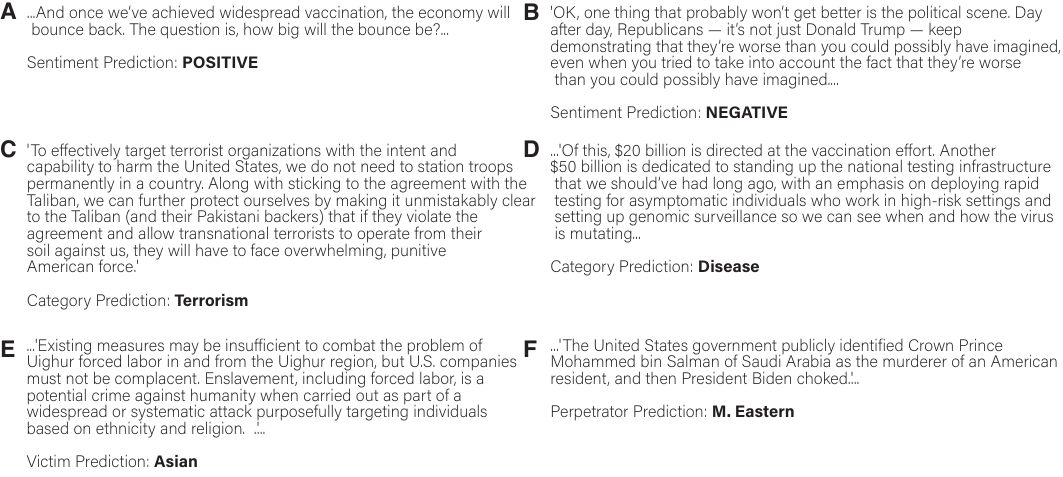}
    \caption{(A, B) Excerpts of articles classified as positive and negative by the sentiment classifier, respectively. (C, D) Excerpts of articles classified as pertaining to Terrorism and Disease by the category classifier, respectively. (E) Excerpt of an article in which the victim is classified as Asian. (F) Excerpt of an article in which the perpetrator is classified as Middle Eastern.}
    \Description{(A, B) Excerpts of articles classified as positive and negative by the sentiment classifier, respectively. (C, D) Excerpts of articles classified as pertaining to Terrorism and Disease by the category classifier, respectively. (E) Excerpt of an article in which the victim is classified as Asian. (F) Excerpt of an article in which the perpetrator is classified as Middle Eastern.}
    \label{fig:text_samples}
\end{figure*}

\end{document}